\documentclass[11pt]{article}
\usepackage[T1]{fontenc}
\usepackage[hmargin=0.88in,vmargin=0.70in]{geometry}
\usepackage{amsmath,amssymb,graphicx,booktabs}
\usepackage{xcolor}
\usepackage[colorlinks=true,allcolors=blue!55!black]{hyperref}
\usepackage{caption,placeins}

\newcommand{\E}{\mathbb E}
\newcommand{\KD}{\mathrm{KD}}
\newcommand{\KL}{D_{\mathrm{KL}}}
\newcommand{\ind}{\mathbf 1}

\title{Generative Ising Models on a Quantum Annealer}

\author{\parbox{0.95\textwidth}{\centering
Carpio, N. A. C.\textsuperscript{1}\thanks{\texttt{ncarpio@estudante.ufscar.br}},
Farias, T. S.\textsuperscript{1,2},
Fernandes, G. P. L. M.\textsuperscript{1,3},\\
Paes, A.\textsuperscript{4},
Villas-Boas, C. J.\textsuperscript{1}\\[0.62em]
{\small
\textsuperscript{1}Department of Computer Science, Federal University of S\~ao Carlos, S\~ao Carlos, SP, Brazil\\
\textsuperscript{2}Department of Physics, University of Innsbruck, Innsbruck, Austria\\
\textsuperscript{3}Wernher von Braun Advanced Research Center, Campinas, SP, Brazil\\
\textsuperscript{4}Department of Information Science, University of North Texas, Denton, TX, USA
}
}}

\date{}

\begin{document}
\maketitle
\begin{abstract}
We distill a configuration-dependent generative energy into an Ising model with fixed pairwise couplings for a quantum annealer. The student contains auxiliary spins whose Boltzmann weights can be summed exactly to obtain a visible free energy. We fit this score to the frozen HyperIsing teacher's energy differences within each class, centered energies, and class probabilities. The visible score supports classification and restoration of MNIST images, and samples from the joint Ising Hamiltonian decode to images conditioned on a requested digit. Across three distilled students, 100 updates using hardware samples increased mean conditional accuracy on the device from 64.39\% to 85.08\% and macro recall in classifier feature space from 19.62\% to 40.47\%. The same refined weights had lower conditional accuracy than the unrefined weights when sampled by simulated quantum annealing on the logical graph. Each evaluation sampler achieved its highest recall when its own samples were used for refinement. These comparisons identify gains that depend on the pairing of refinement and evaluation samplers.
\end{abstract}

\section{Introduction}

Energy-based models assign a scalar energy to each candidate configuration \cite{ebm}. In Boltzmann models, energy differences determine the probability ratios between configurations at fixed temperature \cite{ackley}. This relationship connects inference and generation: one can compare the energies of a specified set of candidates or sample configurations according to their Boltzmann weights. Restricted Boltzmann machines learn energies for binary data by coupling visible spins to hidden spins, and have been used to model handwritten digits \cite{cd}. Summing over the hidden spins can generate effective interactions beyond pairs in the visible energy \cite{decelle}. Neural energy models can also score images directly, as in convolutional image synthesis and completion \cite{xie,du}, or score images and labels jointly using classifier outputs \cite{grathwohl}. Here we ask which preferences of a learned energy can be transferred to a programmable Ising model for inference and generation.

A transverse-field quantum annealer returns binary spin configurations. In a forward anneal, the schedule increases the relative weight of the programmed Ising term with respect to the transverse field. The fields and pairwise couplings defining the Ising term remain fixed for each programmed instance \cite{dwaveanneal}. Benedetti et al.\ used hardware samples to train embedded graphical models and demonstrated generation and classification on handwritten digits \cite{benedetti}. Their subsequent quantum-assisted Helmholtz machine sampled a generative model defined over binary image codes \cite{helmholtz}. Kim, Gyhm, and Park calibrated diabatic annealing to obtain approximate Boltzmann samples for training restricted Boltzmann machines \cite{kimrbm}. In our setting the generative energy is learned before the hardware is used.

A hypernetwork generates parameters for another model \cite{hypernet}. HyperIsing uses a hypernetwork to predict fields and couplings from each spin configuration it scores. This dependence lets a sum of field and pair terms represent higher-order interactions. To transfer the learned energy to a pairwise Hamiltonian with fixed coefficients, we use teacher-student distillation. A trained teacher provides targets that a student with a different parameterization learns to approximate \cite{hinton}. Related work has distilled pretrained diffusion scores into differentiable energies for continuous generation \cite{thornton}. Our student is fitted to energy differences between binary codes within each class, centered teacher energies, and the teacher's class probabilities for a fixed code.

The student contains visible spins that encode the image and auxiliary spins that enrich the effective visible energy. At each code and label, we sum the Boltzmann weights over every auxiliary assignment. At the model's dimensionless reference temperature $T=1$, the visible free energy is the negative logarithm of this sum (Eq.~\eqref{eq:free}). We fit this score to the teacher and program the fixed pairwise Hamiltonian of the joint model on the quantum processing unit (QPU). Figure~\ref{fig:method} illustrates the procedure. On MNIST \cite{mnist}, the visible score supports classification and restoration, and sampled visible codes decode to images of a requested digit.

Finite-run sampling can affect the outputs of learned energy models \cite{nijkamp}. Hardware output can also deviate from the classical Boltzmann law of the final Hamiltonian, with contributions from freeze-out and embedding \cite{freezeout,marshall,marshallq}. The unrefined checkpoint is the set of student weights saved after matching the teacher. We refine separate copies from the same starting weights, using samples either measured on the QPU or generated by simulated quantum annealing (SQA). SQA is a classical path-integral Monte Carlo simulation of a transverse-field model \cite{suzuki,martonak}; in the crossed experiment it runs on the logical student graph without the hardware embedding. We evaluate each set of weights with both samplers. The evaluation measures how often a generated image matches the requested class and how much of that class's reference set is covered in feature space \cite{knn}.

Refining with hardware samples improves both conditional accuracy and recall on the QPU. The same refined weights have lower conditional accuracy than the unrefined weights under logical-graph SQA. For each evaluation sampler, recall is highest after refinement using samples from that sampler.

The Methods section describes distillation, hardware programming, software sampling, and the evaluation protocol. Results and Discussion presents classical inference, crossed refinement and evaluation, and a separate refinement study after embedded-graph SQA pretraining. The appendices provide mathematical derivations, implementation details, and additional experiments.

\section{Model and Methods}

In this section, we define the teacher score and the programmable student, then describe how each sampler supplies examples for refinement. We specify the MNIST implementation and the protocols for classical inference and hardware generation.

\subsection{Binary representation and HyperIsing teacher}

We represent an observation by a binary code $b\in\{0,1\}^{d}$ and use $z=2b-1$ as its spin representation. For images we learn the binary mapping with an autoencoder. We write $D(z)$ for the decoder applied to $(z+1)/2$ and keep both encoder and decoder frozen during all energy-model experiments. We append a spin encoding $\ell(y)$ for the class $y$ to obtain the teacher input $\mathbf{s}=(z,\ell(y))$. A residual multilayer perceptron serves as a hypernetwork that computes the coefficients of the score \cite{hypernet}. The resulting energy is

\begin{equation}
 E_{\mathrm H}(z,y;c)=-\sum_i h_i(\mathbf{s},c)s_i-\sum_{(i,j)\in\mathcal G}J_{ij}(\mathbf{s},c)s_i s_j .
 \label{eq:teacher}
\end{equation}
Here $\mathbf{s}=(z,\ell(y))$ concatenates the latent code $z$ and the class encoding $\ell(y)$, with only the requested class positive. The scalar $c$ specifies the corruption condition supplied to the network. The local field $h_i$ weights spin $s_i$, and the coupling $J_{ij}$ weights the product of spins $s_i$ and $s_j$ for an edge $(i,j)$ of the fixed graph $\mathcal G$. Both coefficient functions are reevaluated for every input to the teacher. During teacher training, ranking favors real codes over corrupted alternatives at the same label and corruption condition. Cross-entropy on the negative energies across labels teaches class discrimination. An energy-magnitude penalty regularizes the score scale. We freeze this teacher and query it at $c=0$ during distillation.

For a fixed class, the hardware accepts one set of fields and pairwise couplings for each anneal. Equation~\eqref{eq:teacher} instead evaluates the coefficient functions at each new configuration. Their dependence on $z$ can produce effective interactions involving three or more spins. A single fixed quadratic program on the teacher's spins cannot in general represent that function. Appendix~\ref{app:representation} gives the precise representation criterion.

\begin{figure}[!htbp]\centering
\includegraphics[width=\linewidth]{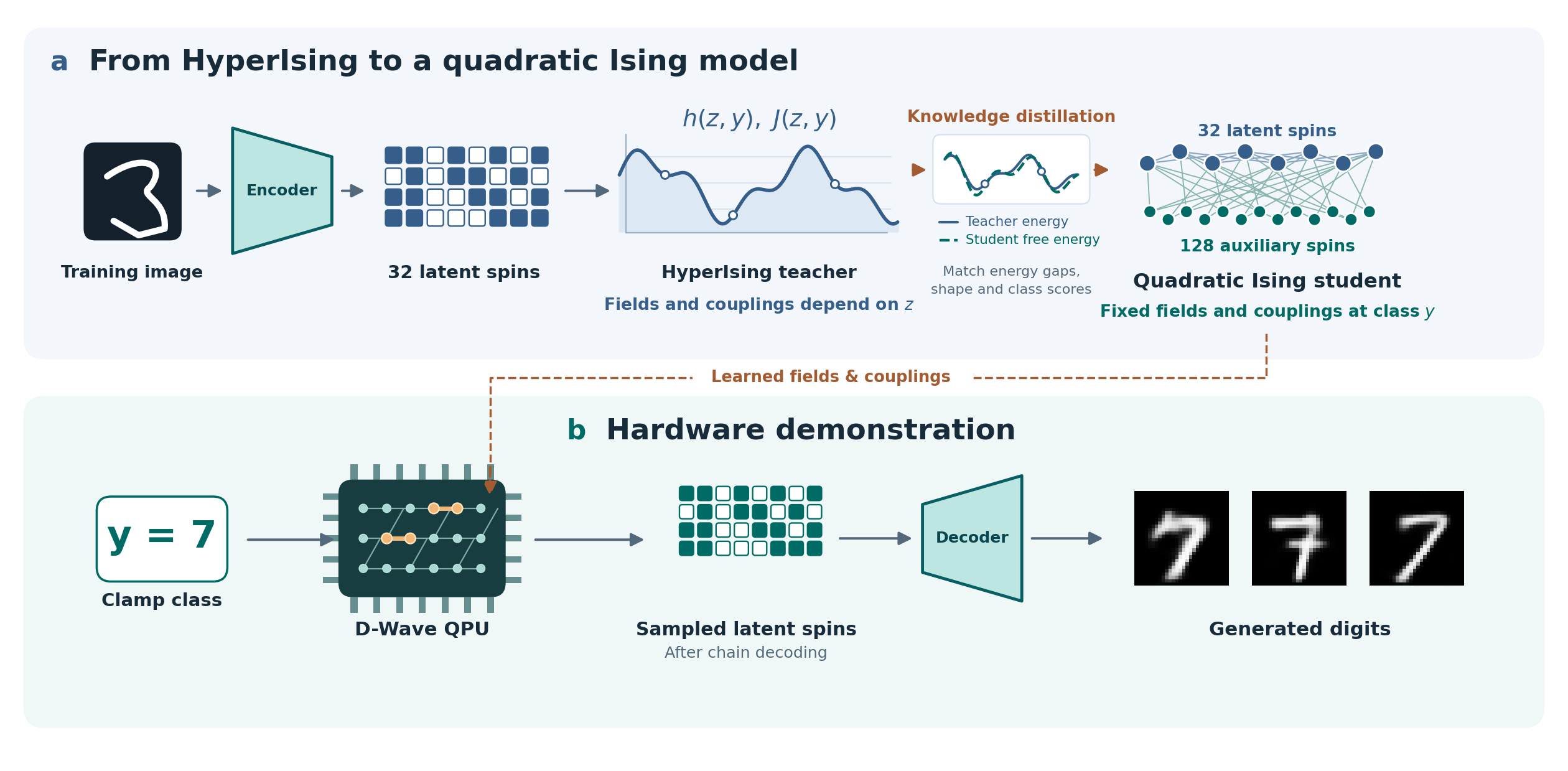}
\caption{\textbf{Energy distillation and hardware generation.} (a) The frozen HyperIsing teacher scores binary image codes. The student learns to match relative teacher energies with its free energy, using a joint Hamiltonian with fixed coefficients at each requested class. The input digit, energy curves and graphs are schematic. (b) With the class clamped, the programmed annealer returns physical reads that are decoded to latent spins and passed through the frozen image decoder. The three sevens are hardware-generated examples, one per acquisition block, from the distilled student after hardware refinement.}
\label{fig:method}
\end{figure}

\subsection{Quadratic student and energy distillation}
At fixed class, the joint energy of the student is
\begin{equation}
 E_\theta(z,a;y)=U_\theta(z;y)-\sum_{j=1}^{m}a_j u_{\theta,j}(z;y),\qquad
 u_{\theta,j}(z;y)=b_j(y)+\sum_i W_{ij}z_i .
 \label{eq:ham}
\end{equation}

Here $\theta$ denotes the trainable coefficients, $z$ contains $d$ visible spins, and $a$ contains $m$ auxiliary spins with $a_j=\pm1$. The term $U_\theta(z;y)$ contains the visible fields, visible-visible couplings and a class-dependent offset. The field $u_{\theta,j}$ combines the bias $b_j(y)$, including the effect of clamping the label, with the contributions of the visible spins coupled to auxiliary $j$. The graph determines which visible-visible couplings and which $W_{ij}$ can be nonzero. The auxiliary spins have no couplings to one another. We use the marginal free energy at logical temperature $T=1$ for distillation:

\begin{equation}
 F_\theta(z;y)=U_\theta(z;y)-\sum_j\log[2\cosh u_{\theta,j}(z;y)] .
 \label{eq:free}
\end{equation}
The first term is the part of the energy independent of the auxiliaries. Each logarithm comes from summing the two Boltzmann weights for $a_j=\pm1$. These sums factorize because the auxiliaries have no couplings to one another at fixed $z,y$ (Appendix~\ref{app:representation}). The resulting $F_\theta(z;y)$ scores a visible code without fixing the auxiliary spins and can encode higher-order visible interactions \cite{decelle}.

We evaluate the frozen teacher at corruption condition $c=0$ for every cached code under all candidate labels. At each distillation step, the student scores a minibatch of code pairs and is fitted to those stored targets. The pairs include codes drawn from data, corrupted codes, uniform codes and teacher sampling trajectories, together with perturbations. We fit the student with the objective
\begin{equation}
 \mathcal L_{\KD} = \mathcal L_\Delta+0.25\,\mathcal L_{\mathrm{shape}}+\mathcal L_{\mathrm{class}}.
 \label{eq:kd}
\end{equation}
The gap loss $\mathcal L_\Delta$ penalizes errors in the difference between the energies of two codes at the same class. The shape loss $\mathcal L_{\mathrm{shape}}$ matches the globally centered energy values, with weight $0.25$. The class loss $\mathcal L_{\mathrm{class}}$ is the Kullback-Leibler divergence from the teacher's soft class probabilities to the student's probabilities for a fixed code. The student learns the coefficients of Eq.~\eqref{eq:ham} by differentiating the loss through Eq.~\eqref{eq:free}, with the teacher fixed. This stage uses the Adam optimizer \cite{adam} and cached teacher targets. Appendix~\ref{app:loss} specifies the objectives and parameters.

\subsection{MNIST implementation}
The image code has $d=32$ spins and the student has $m=128$ auxiliaries. The class encoding has ten entries: $+1$ for the requested class and $-1$ for every other class. The binary autoencoder uses hard quantization with a straight-through gradient \cite{bengioSTE}; Appendix~\ref{app:binary-ae} specifies the architecture and training. Dense connects every pair of visible spins and every visible spin to every auxiliary. Sparse~8/32 gives each visible spin eight visible neighbors and retains 32 visible neighbors per auxiliary. Sparse~8/16 gives each visible spin eight visible neighbors and each auxiliary 16 visible neighbors. The main experiments use Dense; Appendix~\ref{app:architecture} compares the three variants. The common additive constant $128\log2$ in $U_\theta$ cancels in the distillation and refinement objectives (Appendix~\ref{app:representation}).

\subsection{Classical inference}
For classification we evaluate every candidate class at the same image code and choose the label with the lowest energy. Restoration searches over image codes and labels using an energy augmented by a pixel-fidelity term on the observed pixels. The visible-space procedure uses $F_\theta$ and the joint procedure uses $E_\theta$ with explicit auxiliary spins. Appendix~\ref{app:inference} gives the initialization, fidelity term, sweep schedule, and comparison protocol.

\subsection{Programming and reading the annealer}
\label{sec:qa}
\begin{samepage}
After fixing the requested class, we map the student's joint Ising energy in Eq.~\eqref{eq:ham} to the physical graph. In the device sign convention the programmed energy is \cite{dwaveanneal}
\begin{equation}
 H_{\rm prog}(x)=\sum_q h_q^{\rm p}x_q+
 \sum_{(q,r)\in\mathcal E_{\rm p}}J_{qr}^{\rm p}x_qx_r .
 \label{eq:programmed_ising}
\end{equation}
Here $x_q=\pm1$ is the spin value of physical qubit $q$ and $\mathcal E_{\rm p}$ is the set of physical couplers used by the embedding. The dimensionless $h_q^{\rm p}$ is the local bias at $q$, and $J_{qr}^{\rm p}$ is the pairwise coupling on edge $(q,r)$; a negative coupling favors aligned spins. A minor embedding represents each logical variable by a connected group of qubits. The group's internal ferromagnetic couplings favor a shared value, and the contacts between groups carry the logical pairwise interactions \cite{choi}. The class is fixed through the programmed coefficients and is not measured as a separate logical variable.
\end{samepage}

\begin{samepage}
The forward anneal is governed by the physical Hamiltonian \cite{dwaveanneal}
\begin{equation}
 \frac{\widehat H_{\rm p}(s)}{h_{\rm P}}
 =-\frac{A(s)}{2}\sum_q X_q+
 \frac{B(s)}{2}\,H_{\rm prog}(\{Z_q\}).
 \label{eq:physical_anneal}
\end{equation}
Here $h_{\rm P}$ is Planck's constant, so the equation is expressed in frequency units, and $s$ is the normalized annealing coordinate from 0 to 1. The schedules $A(s)$ and $B(s)$ set the strengths of the transverse field and the programmed Ising term, respectively. The Pauli operator $X_q$ flips spin $q$ in the measurement basis. The operator $Z_q$ is diagonal in that basis with eigenvalues $\pm1$; $H_{\rm prog}(\{Z_q\})$ is Eq.~\eqref{eq:programmed_ising} with each $x_q$ replaced by $Z_q$. The transverse term initially dominates and connects configurations that differ by a single spin flip. As its relative strength decreases, the programmed Ising term dominates. The local biases and couplers remain fixed during this change in the device schedules. The final readout returns one value per physical qubit.
\end{samepage}

We reuse a fixed minor embedding of 160 logical spins into 1,071 physical qubits on Advantage2\_system1. The coefficient scale is $0.75563$ and the ferromagnetic chain magnitude is set to $1.9$ in programmed units, with automatic rescaling disabled. The forward schedule is linear and lasts $500\,\mu\mathrm{s}$, with 100 reads per request. The spin-reversal transformation used before submission is \cite{dwavegauge}

\begin{equation}
 \widetilde h_q^{\rm p}=g_qh_q^{\rm p},\qquad
 \widetilde J_{qr}^{\rm p}=g_qg_rJ_{qr}^{\rm p},\qquad
 \widetilde x_q=g_qx_q .
 \label{eq:gauge}
\end{equation}
Here $g_q=\pm1$ is the gauge sign assigned to physical qubit $q$, and tildes denote the transformed coefficients and spin values. Applying the three transformations together preserves each field and coupling contribution to Eq.~\eqref{eq:programmed_ising}. We transform each hardware read back to its original gauge and recover the logical variables by
\begin{equation}
 \widehat v_i=\operatorname{sgn}_{+}\!\left(\sum_{q\in C_i}g_q\widetilde x_q\right).
 \label{eq:decoding}
\end{equation}

The set $C_i$ contains the physical qubits representing logical variable $i$. A chain is broken if its physical spins disagree after undoing the gauge. Equation~\eqref{eq:decoding} assigns one logical value by majority vote, with ties assigned to $+1$. Appendix~\ref{app:rrs} compares this rule with restricted resampling \cite{marshall}. The decoded vector $\widehat v=(\widehat z,\widehat a)$ contains 32 image-code spins $\widehat z$ and 128 auxiliary spins $\widehat a$. Each read gives one image $D(\widehat z)$ (Fig.~\ref{fig:method}b). During refinement, $\widehat z$ also supplies a negative code for the free-energy gradient, which sums over auxiliary states analytically.

\subsection{Simulated quantum annealing}

Both software samplers use a Suzuki-Trotter representation \cite{suzuki} and path-integral Monte Carlo \cite{martonak}. Logical-graph SQA operates on the unembedded 160-spin student graph. It uses 16 Trotter slices, 64 sweeps, logical $T=1$, and a transverse field decreasing from 4 to 0.05. Embedded-graph SQA simulates the programmed physical graph, including the ferromagnetic chains, using the published Advantage2\_system1 schedules $A(s)$ and $B(s)$ \cite{dwaveschedules}. The embedded-graph initialization uses a simulation bath parameter of $20$\,mK, 32 Trotter slices and 64 sweeps. Appendix~\ref{app:sampling} details both protocols. In all refinement experiments the update uses the visible free energy at the logical reference temperature.

\subsection{Sample-based refinement}
For each update, we differentiate the following difference of batch means and pass the gradient to Adam:
\begin{equation}
 \widehat{\mathcal L}_{\mathrm{ref}}
 =\frac{1}{|D_y|}\sum_{z\in D_y} F_\theta(z;y)
  -\frac{1}{|M_y|}\sum_{z\in M_y}F_\theta(z;y),
 \label{eq:surrogate}
\end{equation}
In both sums, $F_\theta(z;y)$ is the student free energy defined in Eq.~\eqref{eq:free}. The positive batch $D_y$ contains codes of real images with label $y$. The negative batch $M_y$ contains the visible components returned by the selected sampler at the current student weights and the same label. The vertical bars give the batch sizes, counting repeated samples. Holding both batches fixed, the update favors lower mean free energy for the data relative to the sampled codes. It estimates the negative log-likelihood gradient at the reference temperature when the sampler follows the model distribution at that temperature; otherwise it is a sampler-dependent update (Appendix~\ref{app:sampling}). Each method refines the same starting student for 100 updates with one class per update, using ten shuffled passes through the classes, with 400 real codes and 100 sampled codes per update. The optimizer and class order are matched across QPU and logical-graph SQA refinement within each seed.

\subsection{Evaluation}
Conditional accuracy is the fraction of generated images assigned to the requested label by a frozen convolutional neural network (CNN). Our coverage metric is class-conditional nearest-neighbor recall: the fraction of reference images lying in the generated set's neighborhoods in CNN feature space \cite{knn}. We compute recall within each requested digit and average over digits for macro recall. We also measure the fraction of distinct sampled latent codes and the Fr\'echet distance using the CNN features \cite{fid}. Appendix~\ref{app:metrics} defines the metrics and reference sets. In the crossed experiment we average sampling blocks or streams within each trained checkpoint before computing the mean and sample standard deviation over three training seeds. In each single-trajectory experiment, hardware error bars describe the spread across acquisition blocks at fixed weights.

\section{Results and Discussion}
\subsection{Model size and hardware footprint}
\begin{figure}[!htbp]\centering
\includegraphics[width=\linewidth]{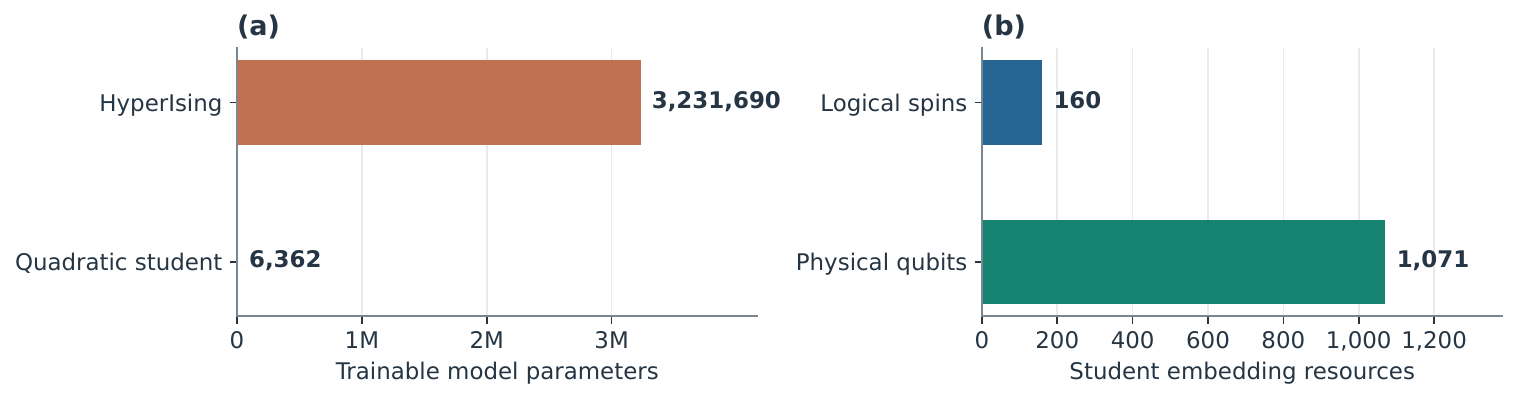}
\caption{\textbf{Model size and hardware footprint.} (a) Trainable coefficients of the teacher and dense student, excluding the shared autoencoder and evaluation classifier. (b) Logical spin count of the class-clamped student and physical qubits used by its embedding in the main hardware experiments.}
\label{fig:resources}
\end{figure}

The dense student has 6,362 trainable coefficients, compared with 3,231,690 for the teacher. With the class fixed, the 160 remaining logical spins occupy 1,071 physical qubits in the embedding used for the main hardware experiments (Fig.~\ref{fig:resources}). Appendix~\ref{app:representation} enumerates the student coefficients. Appendix~\ref{app:architecture} compares embeddings for different student graphs.

\subsection{Classification and image restoration}
\begin{figure}[!t]\centering
\includegraphics[width=.94\linewidth]{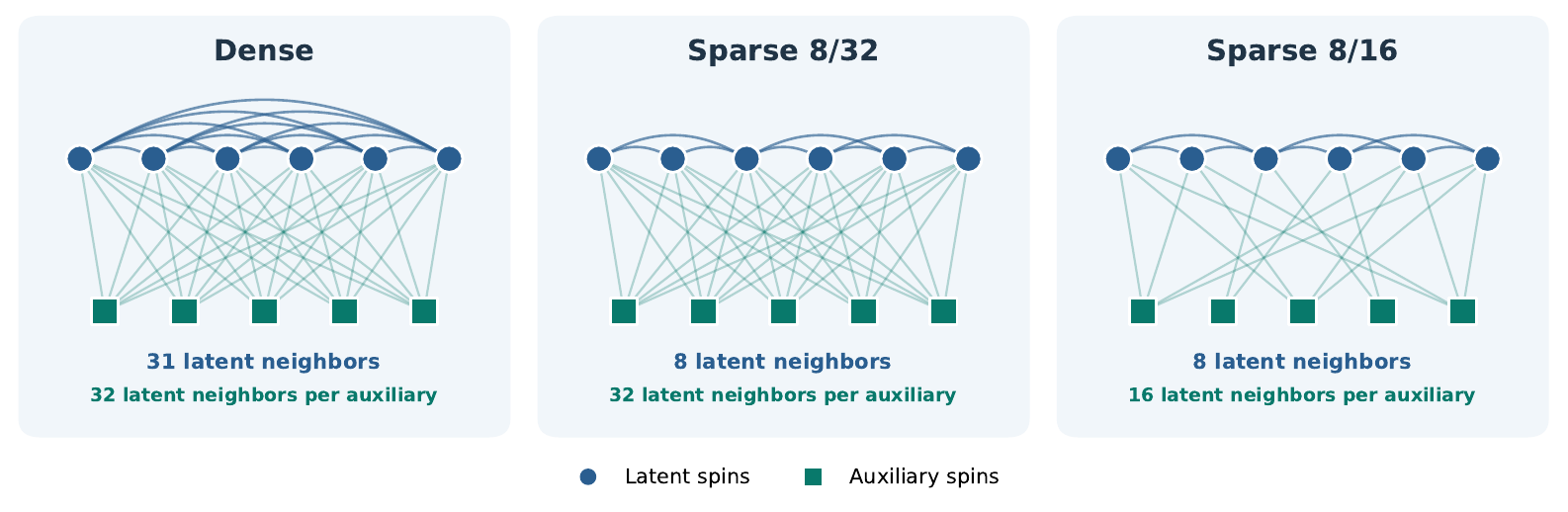}\par\smallskip
\includegraphics[width=.94\linewidth]{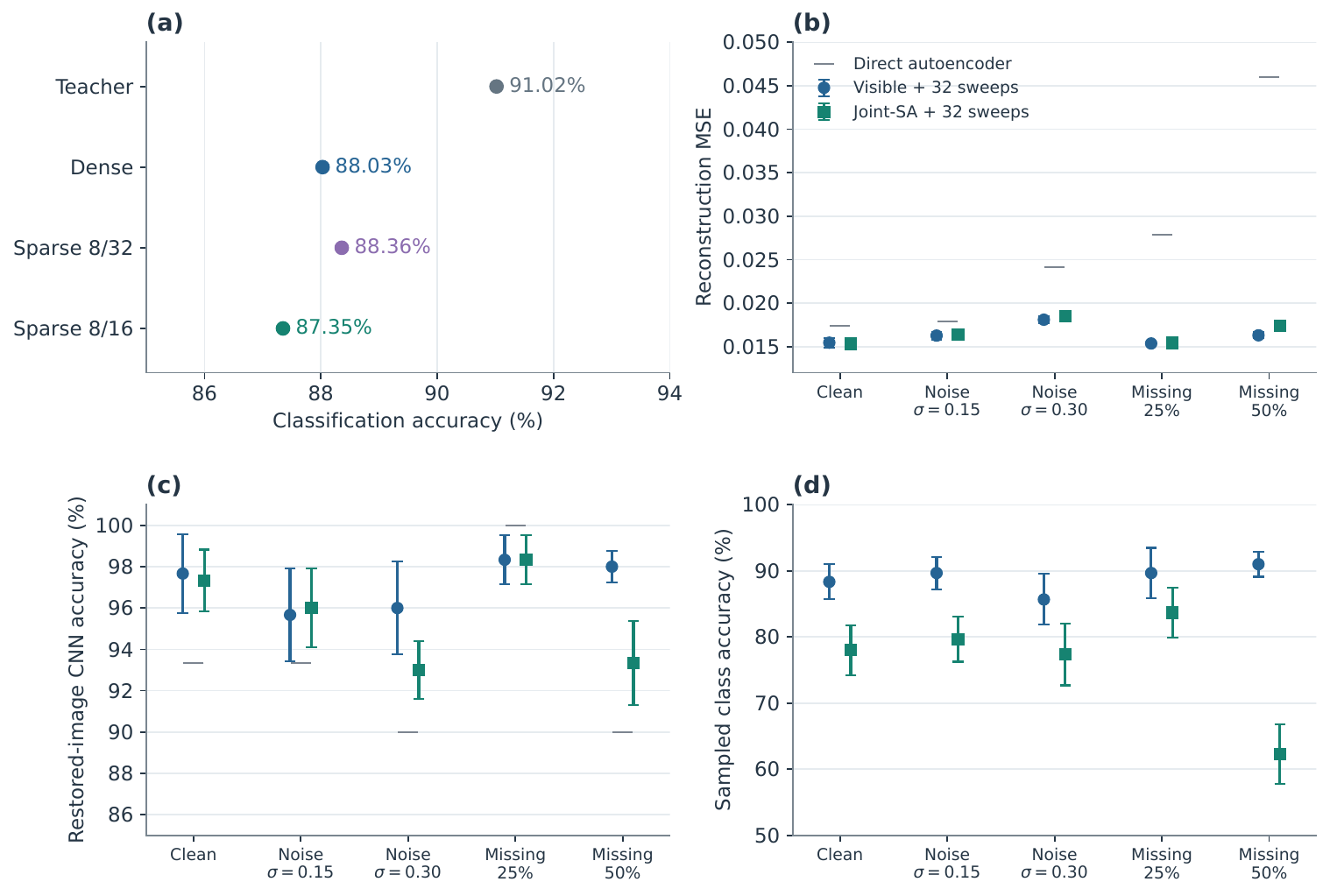}
\caption{\textbf{Student architectures and classical inference.} Top: reduced schematics for the three graphs, with circles for latents and squares for auxiliaries. The labels specify degrees of the full graph. (a) Energy-based classification on 10,000 MNIST test images, one checkpoint per model. (b-d) Dense-student restoration for clean, noisy and masked inputs. Metrics report mean squared error to the clean image, accuracy of the fixed CNN on the restored image, and accuracy of the sampled label, respectively. In (b-d), the visible and joint sampling results are means and sample standard deviations over five sampler seeds using the same 30 images and two trajectories per image. Direct autoencoder results are shown where an image-level baseline is defined.}
\label{fig:inference}
\end{figure}

The student classifies a fixed code by choosing the label with minimum free energy. On the full MNIST test set, the teacher classified $91.02\%$ correctly, and the dense, Sparse~8/32, and Sparse~8/16 students reached $88.03\%$, $88.36\%$, and $87.35\%$, respectively (Fig.~\ref{fig:inference}a). The sparse variants use visible degree eight and respectively 32 or 16 visible neighbors per auxiliary spin (top of Fig.~\ref{fig:inference}). This comparison used one archived checkpoint per architecture. These checkpoints are also used in the architecture comparison in Appendix~\ref{app:architecture}; the main hardware refinement experiments start from separate Dense checkpoints.

For restoration an archived dense checkpoint also used for Fig.~\ref{fig:inference}a samples both the latent code and its label using the learned energy with a pixel-fidelity term on the observed pixels. Visible sampling proposes flips only in the 32 latent spins, with the auxiliaries already summed out in $F_\theta$. Joint simulated annealing (Joint SA) explicitly updates the 32 latent and 128 auxiliary spins using $E_\theta$. Both update the class and run 32 annealing sweeps followed by 32 additional sweeps at the final temperature (the ``+32'' in the figure). The direct autoencoder applied to the corrupted image supplies an image-level baseline. Both complete sampling procedures gave lower reconstruction mean squared error than a direct autoencoder pass across all five observation conditions (Fig.~\ref{fig:inference}b). With 50\% of pixels masked, visible-space restoration achieved $98.0\%$ CNN accuracy and $91.0\%$ sampled-label accuracy; the joint method gave $93.3\%$ and $62.3\%$, respectively. These accuracies report image identity and label inference separately. The direct pass is a reference for the complete procedure; a comparison isolating the learned energy would use an equal search budget for a fidelity-only objective. Appendix~\ref{app:inference} specifies the fidelity term, observations, and evaluation.

The sparse variants retain classification performance close to Dense in the archived checkpoints. This comparison probes label preference after encoding, because the same visible code is scored under each possible class. Restoration probes another use of the learned score: multiple codes can agree with the observed pixels, and the objective uses the learned energy to rank them. The visible sampler evaluates this score after summing over the auxiliaries, whereas joint annealing must explore those spins alongside the code. Under the reported sweep schedule the visible sampler gives stronger label recovery in the half-masked case. This comparison is between two finite-sweep procedures: they use different visible targets during cooling and the same nominal target at the final temperature (Appendix~\ref{app:inference}). Reducing the visible degree alone still leaves the dense visible-auxiliary block, so the physical footprint also depends on how those couplings embed on the hardware (Fig.~\ref{fig:resources} and Appendix~\ref{app:architecture}).

\subsection{Learning Dynamics During Hardware Refinement}
\begin{figure}[!htbp]\centering
\includegraphics[width=\linewidth]{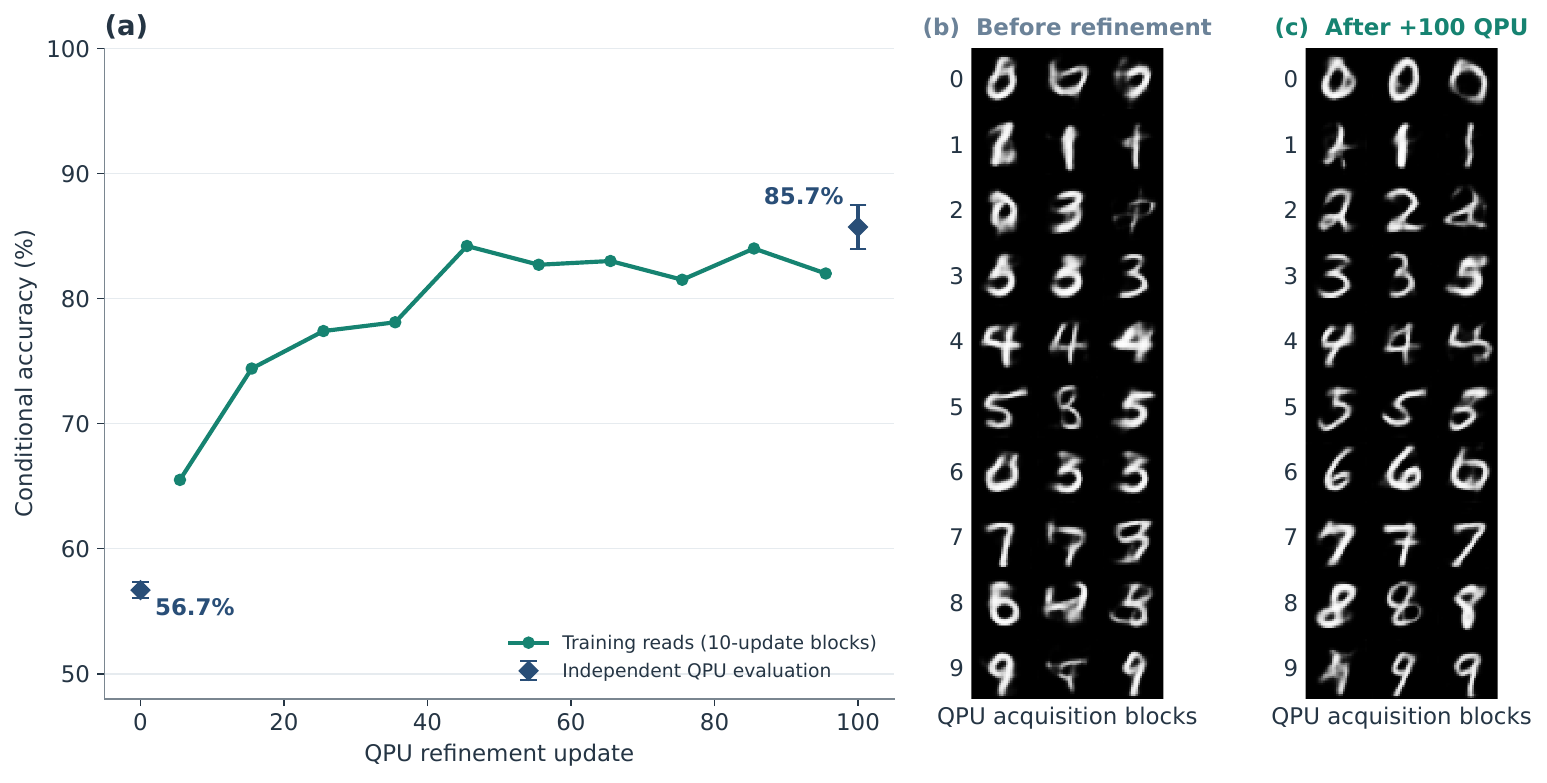}
\caption{\textbf{Accuracy during hardware refinement.} (a) Circles show one accuracy per ten-update block of training reads, taken before the corresponding parameter updates. Each block pools 100 reads per requested class. Diamonds show the mean and sample standard deviation of independent hardware evaluations across three acquisition blocks at the frozen and final checkpoints. (b,c) QPU-generated digits at the two checkpoints. Each column corresponds to one acquisition block. The same read positions within a block and class are used before and after refinement, based on an archived random index per class and block. Images are decoded directly after majority-vote chain decoding.}
\label{fig:trajectory}
\end{figure}

Figure~\ref{fig:trajectory} follows one distilled student through 100 updates. We evaluated the images decoded from the saved training reads, grouping ten consecutive requests so every block includes all requested digits. The accuracy of these reads rose from $65.5\%$ over updates 1-10 to $84.2\%$ over updates 41-50 and ranged from $81.5\%$ to $84.0\%$ over the remaining blocks. Independent evaluations at the two fixed checkpoints gave $56.68\pm0.65\%$ before and $85.72\pm1.79\%$ after refinement. Both fixed checkpoints were evaluated after training, using 6,000 QPU reads per endpoint. Appendix~\ref{app:surrogate} shows the logged training objective for the same trajectory. The training-read curve tracks the model as its parameters change; the independent reads at fixed endpoints test whether the improvement carries over to generation after training. The two measurements serve different purposes: the first shows the progression of the update, and the second compares the two generators under a common readout protocol.

\subsection{Crossed Evaluation of Refinement Samplers}
\begin{figure}[!t]\centering
\includegraphics[width=.94\linewidth,height=.71\textheight,keepaspectratio]{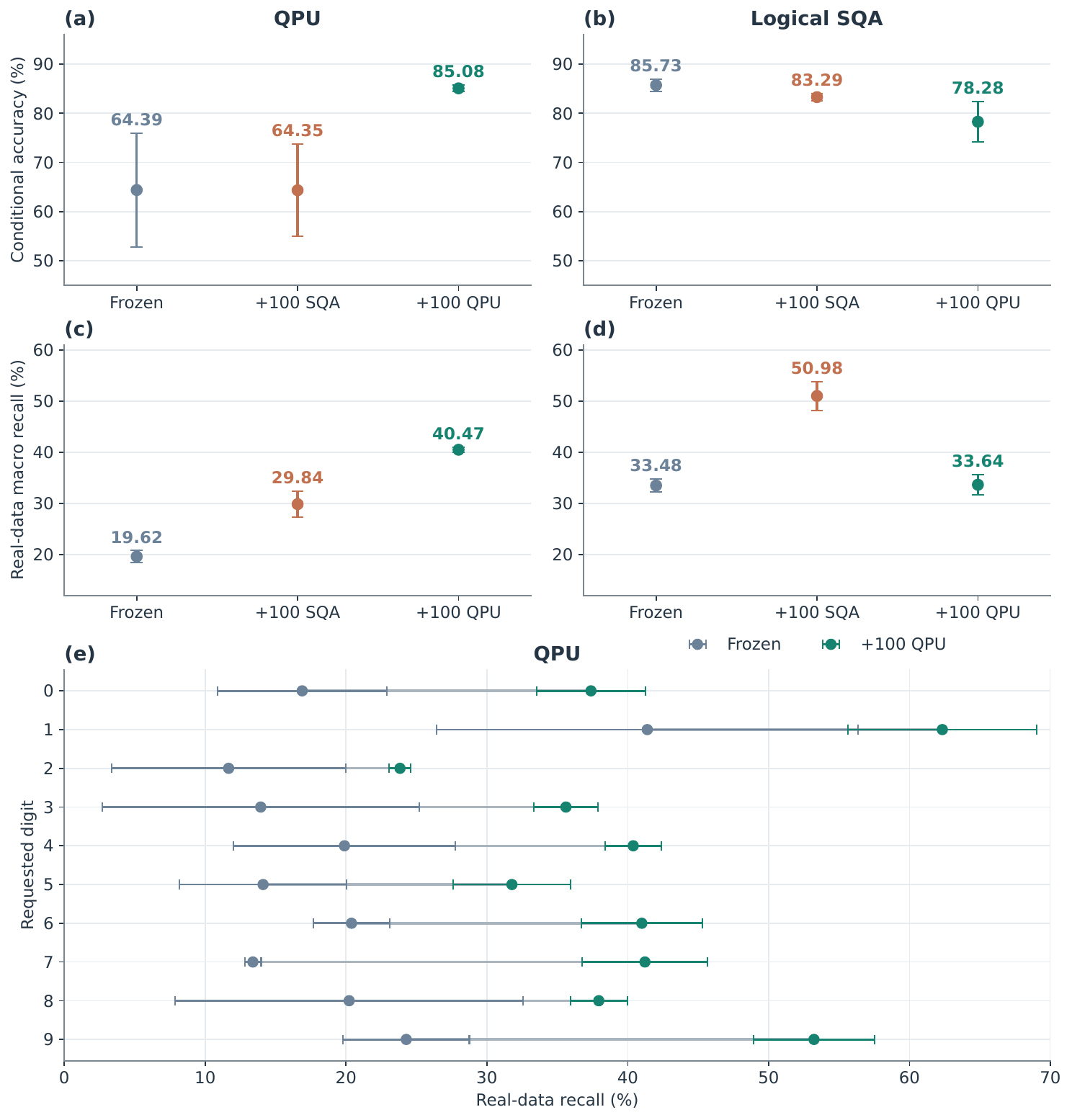}
\caption{\textbf{Crossed evaluation and coverage by requested digit.} (a-d) Frozen distilled students and their matched refinements with logical-graph SQA or QPU negatives. Columns identify the sampler used for evaluation. (e) Real-image recall by class under QPU evaluation for frozen and hardware-refined students. Points and bars denote the mean and sample standard deviation over three training seeds after averaging acquisition blocks or simulation streams within each seed. The trajectory in Fig.~\ref{fig:trajectory} is one of the three seeds in this comparison.}
\label{fig:cross}
\end{figure}

We evaluated three distilled students and their two refined variants with both the QPU and logical-graph SQA (Fig.~\ref{fig:cross}). Appendix~\ref{app:paired} reports each within-seed change. Under QPU evaluation, using hardware reads for refinement raised mean conditional accuracy from $64.39\%$ to $85.08\%$ and macro recall from $19.62\%$ to $40.47\%$. The paired changes across seeds were $20.68\pm11.60$ and $20.86\pm1.45$ percentage points, respectively. Using logical-graph SQA reads for refinement gave $64.35\%$ QPU accuracy and $29.84\%$ QPU recall.

With logical-graph SQA evaluation, the hardware-refined weights gave $78.28\%$ accuracy and $33.64\%$ recall, compared with $85.73\%$ and $33.48\%$ for the frozen weights. Logical-graph SQA refinement gave $83.29\%$ accuracy and $50.98\%$ recall in that evaluator. The change in generation quality depends on the pairing of refinement and evaluation samplers. Under QPU evaluation the hardware-refined model improved real-image recall for every digit in each of the three seeds. The largest mean increases were for digits 9 and 7, at 28.94 and 27.83 percentage points, respectively.

Replacing the source of negative examples changes which configurations contribute to Eq.~\eqref{eq:surrogate}. The crossed design holds the evaluation sampler fixed when comparing these updates. The best recall for each evaluation sampler follows refinement with the corresponding sampler, consistent with adaptation to the configurations it delivers. Conditional accuracy measures whether generated images are recognizable as the requested class, while recall measures how much of the reference feature space they cover. Their different responses to refinement show why both are needed to characterize generation. In particular, the logical-graph SQA-refined checkpoint improves recall when evaluated on the QPU with almost unchanged conditional accuracy. That pattern indicates a change in the spread of outputs even when the fraction assigned to the requested class is similar. Conversely, the hardware-refined checkpoints improve both measures on the QPU but give lower accuracy under logical-graph SQA. A gain measured with hardware readout therefore need not transfer to a different evaluator applied to the same weights.

Calibrated diabatic annealing illustrates a complementary strategy: tune the hardware to produce samples compatible with a target Boltzmann law \cite{kimrbm}. Here the annealing protocol is held fixed, and the model coefficients are updated using the samples it delivers. The hardware sampling pipeline provides several opportunities for a mismatch between the logical and physical samples. Annealers can exhibit instance-dependent freeze-out behavior \cite{freezeout}; minor embedding changes the system from which samples are drawn \cite{marshallq}. In the separate frozen-model experiment, resampling broken chains changes the class accuracy obtained from the same physical reads, and subsequent classical sweeps bring the two decoding choices to similar accuracies (Appendix~\ref{app:rrs}; see Ref.~\cite{marshall}). The crossed design measures the effect of changing the sampler that supplies the negative batch under the tested protocols. It does not isolate which aspect of hardware sampling accounts for the difference. A paired experiment that changes one element of the hardware protocol at a time, holding initialization, data batches and read budget fixed, could test whether refinement transfers across annealing schedules or embeddings.

\subsection{Hardware Refinement after Embedded-Graph SQA}
\begin{figure}[!tb]\centering
\includegraphics[width=.94\linewidth,height=.66\textheight,keepaspectratio]{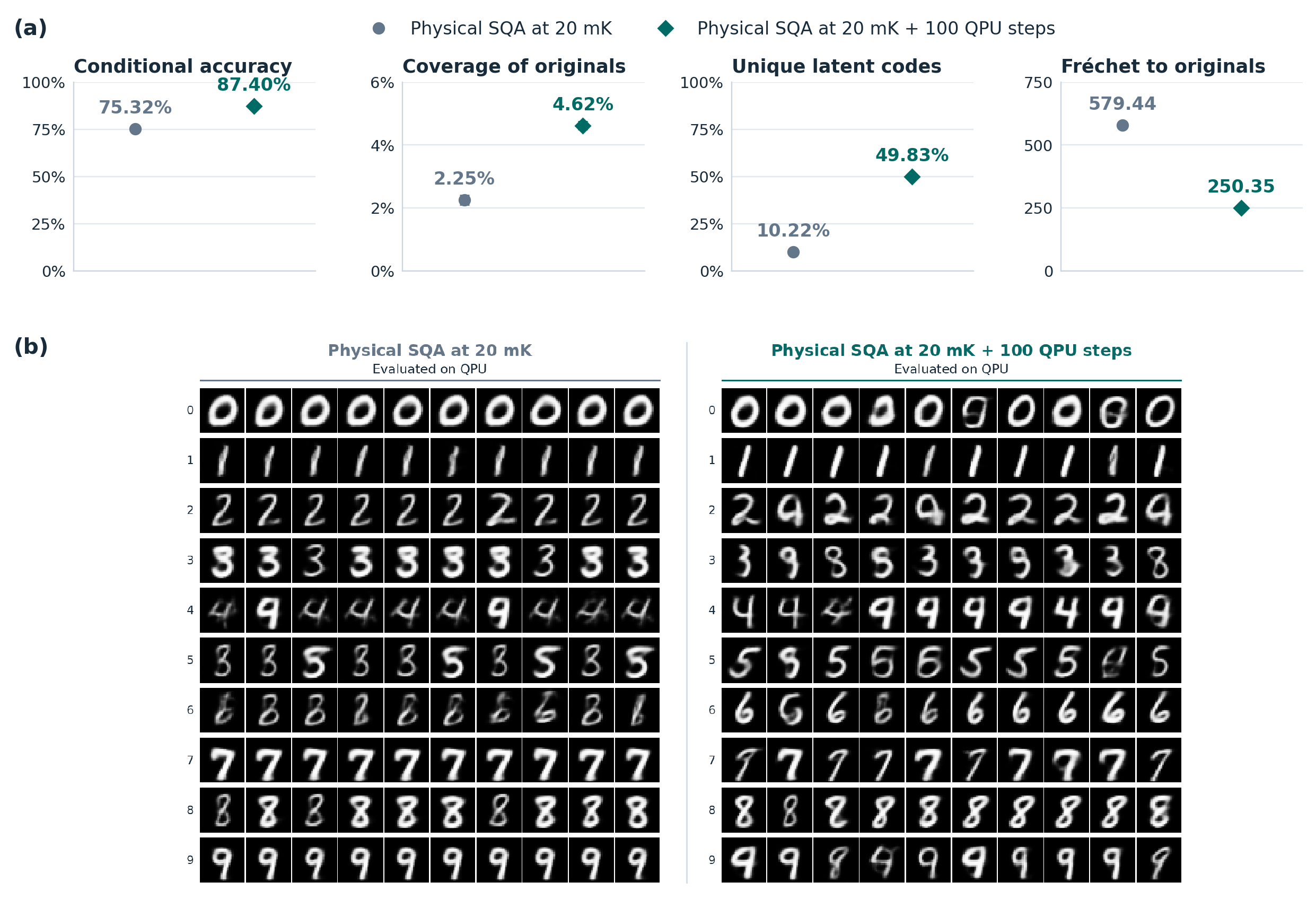}
\caption{\textbf{Generation after embedded-graph SQA initialization and hardware refinement.} (a) QPU evaluation of the same student before and after 100 hardware-sampled updates. Coverage denotes real-image macro recall; the Fr\'echet score uses frozen CNN features. The reported scores are means over three acquisition blocks. (b) First ten reads per requested digit from the same designated block and gauge for each checkpoint, decoded with the same autoencoder.}
\label{fig:physical}
\end{figure}

We trained a separate student from random coefficients for 1,000 Adam updates using negatives from embedded-graph SQA at 20\,mK, then applied 100 hardware-sampled updates. Both checkpoints were evaluated on the QPU under the main hardware protocol. Conditional accuracy increased from $75.32\pm0.98\%$ to $87.40\pm0.84\%$, with a paired by-block change of $12.08\pm1.55$ percentage points. The fraction of distinct latent codes increased from $10.22\%$ to $49.83\%$, and mean real-image recall increased from $2.25\%$ to $4.62\%$. The corresponding CNN-feature Fr\'echet score decreased from 579.44 to 250.35 (Fig.~\ref{fig:physical}). Recall relative to autoencoder reconstructions increased from $8.80\%$ to $20.57\%$.

We sum the QPU access times returned for each request \cite{dwtime}. The 100 refinement requests totaled 10.05\,s of QPU access; including both endpoint evaluations, the total was 22.11\,s. Each endpoint used 6,000 reads in three acquisition blocks. The baseline and final acquisitions took place in separate time windows before and after refinement, respectively; the paired changes summarize these acquisitions.

In this trajectory, high conditional accuracy coexisted with low real-image recall. The post-refinement reads contained more distinct codes and covered a larger fraction of the reconstruction reference set than the earlier reads. The two training routes therefore address distinct starting points: distillation transfers a frozen teacher's preferences, while embedded SQA pretrains the student with simulated samples. Both can be followed by the same hardware refinement procedure. The embedded-graph SQA comparison follows a single checkpoint trajectory across two acquisition windows, so it reports a within-trajectory change. The crossed experiment instead changes the refinement sampler while matching the starting checkpoint, training protocol, and evaluation sampler within each comparison.

\section{Conclusion}
Distillation transfers selected energy differences and class preferences from a configuration-dependent HyperIsing teacher to a fixed pairwise Ising model. The student represents two forms of the same model. Its visible free energy sums over the auxiliary spins and can capture effective interactions beyond pairs. Its joint energy retains the fixed fields and couplings used to program the annealer. In our MNIST experiments, the visible free energy supports classification and restoration. The annealer supplies class-conditioned spin configurations that the same frozen decoder turns into images.

The crossed experiment shows that the gains from refinement depend on the sampler used for evaluation. Refinement with hardware samples improves both conditional accuracy and coverage on the QPU, while the same refined weights yield lower accuracy under logical-graph SQA. For both samplers, coverage in classifier feature space is highest when the same sampler supplies the negative examples during refinement. This pattern is consistent with adaptation to the sampling protocol; the experiment does not identify the physical origin of the differences. The separate student initialized with embedded-graph SQA also improves after hardware refinement, but its real-image recall remains low despite high conditional accuracy. These results motivate measuring both fidelity to the requested class and coverage when assessing hardware generation, and testing whether the gains transfer across embeddings or schedules.

\section*{Acknowledgments}

This work was supported by Petrobr\'as/CENPES, grant No.~2025/00505-2 and by the S\~ao Paulo Research Foundation (FAPESP), under Grants Nos.~2025/27502-3, 2023/15739-3, 2022/00209-6, and 2025/15490-0. C. J. V.-B. acknowledges the support from the National Council for Scientific and Technological Development (CNPq), grants No. 422316/2023-7 and No. 302234/2026-8.

\FloatBarrier
\bibliographystyle{unsrt}
\bibliography{references}

\clearpage

\appendix
\setcounter{figure}{0}
\renewcommand{\thefigure}{A\arabic{figure}}
\renewcommand{\theHfigure}{appendix.\arabic{figure}}
\renewcommand{\thesection}{\Alph{section}}
\section{Representation by a Fixed Ising Hamiltonian}
\label{app:representation}
At fixed class and corruption condition, any score on the binary latent space admits the expansion
\begin{equation}
 E(z)=\sum_{S\subseteq\{1,\ldots,d\}}\widehat E(S)\chi_S(z),\qquad
 \chi_S(z)=\prod_{i\in S}z_i,\qquad
 \widehat E(S)=2^{-d}\sum_{z\in\{-1,1\}^d}E(z)\chi_S(z).
\end{equation}
Here $d$ is the number of visible spins, $S$ is a subset of spin indices, and $\chi_S$ is the product of those spins, with $\chi_\varnothing=1$. The coefficient $\widehat E(S)$ is the uniform average of the score multiplied by that product. To see why the expansion is unique, take two subsets $S$ and $R$. Their basis functions multiply to $\chi_{S\triangle R}$, where $S\triangle R$ is the set of indices belonging to exactly one of them. Its uniform average vanishes unless $S=R$. The $2^d$ functions therefore form an orthonormal basis for the scores on the $2^d$ configurations. A fixed quadratic Ising energy on the same $d$ spins has coefficients only for $|S|\le2$. Conversely those coefficients define the constant, fields and couplings. Thus a single fixed quadratic energy represents $E$ exactly on the same spins if and only if $\widehat E(S)=0$ for every $|S|>2$. For example, setting $J_{12}(z)=z_3$ creates the energy term $-z_1z_2z_3$. Configuration-dependent fields and couplings permit this behavior even though each evaluation uses a sum of field and pairwise contributions.

For the student, summing over the independent auxiliary spins gives
\begin{equation}
 \sum_a \exp[-E_\theta(z,a;y)/T]
 =\exp[-U_\theta(z;y)/T]\prod_j 2\cosh[u_{\theta,j}(z;y)/T].
\end{equation}
The sum on the left runs over all assignments of $a$ with $z,y$ fixed, and $T$ is the temperature in logical energy units. The product on the right has one factor per auxiliary spin; each factor is the sum of that spin's two weights. The marginal free energy is $-T$ times the logarithm of the sum, giving Eq.~\eqref{eq:free} at $T=1$. At $T=1$, the implementation constant $128\log 2$ normalizes the auxiliary contribution to zero when all auxiliary fields vanish. Gap, centered, class-probability and phase-difference calculations are invariant to this constant. The dense model counts are 32 visible biases, 128 auxiliary biases, 10 label biases, 496 visible couplings, 4,096 visible-auxiliary couplings, and 1,600 label-spin couplings across the visible and auxiliary spins. Their sum is 6,362. After clamping the class, 4,592 pairwise interactions remain potentially nonzero on the 160 logical spins.

\section{Teacher and Student Training Objectives}
\label{app:loss}
The teacher hypernetwork has widths $256,128,64$, two residual MLP blocks per level, a scalar-to-16-dimensional condition embedding, and separate heads for fields and edges. The sparse teacher graph has visible degree eight and label-visible edges. We use the archived sparse8 compact checkpoint trained with the ranking and cross-entropy objective (seed 0, step 4,000). The pairwise ranking penalty for the teacher is
\begin{equation}
 \ell_{\mathrm{rank}} = \log\!\bigl(1+\exp[E_+-E_-]\bigr).
\end{equation}
Here $E_+=E_{\mathrm H}(z_+,y;c)$ is the score of the real code and $E_-=E_{\mathrm H}(z_-,y;c)$ is the score of its corrupted alternative, at the same label and condition. The penalty decreases when the real code receives lower energy. It is averaged over pairs with different codes; the alternatives are constructed using random latent flips and uniform codes. For classification, a softmax normalizes the exponentials of the ten negative clean-code energies, giving higher probability to labels with lower energy. The teacher loss adds cross-entropy against the true label and $10^{-4}$ times the mean squared energy over real and corrupted codes.

The three terms in the distillation objective, Eq.~\eqref{eq:kd}, are
\begin{align}
 \mathcal L_\Delta &=
 \left\langle
 \frac{[F_\theta(z';y)-F_\theta(z;y)-E(z',y)+E(z,y)]^2}{\sigma^2}
 \right\rangle_{z,z',y},\\
 \mathcal L_{\rm shape} &=
 \left\langle\frac{[(F_\theta-\overline F_\theta)-(E-\overline E)]^2}{\sigma^2}\right\rangle_{\mathcal B},\\
 \mathcal L_{\rm class} &=
 \left\langle \KL\!\left(
 \operatorname{softmax}_y[-E(z,y)]\,\middle\|\,
 \operatorname{softmax}_y[-F_\theta(z;y)]\right)\right\rangle_{z}.
\end{align}
In these expressions $E(z,y)=E_{\mathrm H}(z,y;0)$ is the frozen teacher's score evaluated at $c=0$, while $F_\theta(z;y)$ is the student's free energy. A pair $(z,z')$ is compared at the same label $y$ in $\mathcal L_\Delta$. The scale $\sigma=\max(0.1,\mathrm{SD}_{\rm train}(E))$ is the standard deviation of cached training teacher scores, floored at $0.1$; the two squared errors are divided by $\sigma^2$. The brackets denote averages: the gap loss uses pairs and labels, the shape loss uses the set $\mathcal B$ of all code-label entries in the minibatch, and the class loss uses both codes from each pair. The bars $\overline F_\theta$ and $\overline E$ denote the separate means of student and teacher scores over $\mathcal B$. In the class loss, $\operatorname{softmax}_y$ normalizes over labels at fixed code, and $\KL(P\|Q)$ compares the teacher distribution $P$ to the student distribution $Q$ in that order. The three dense starting checkpoints in the crossed experiment use 9,000 distillation updates, each with 128 pairs drawn with replacement from 12,000 cached training pairs, with $\sigma=6.59875$. The Adam learning rate is $10^{-3}$, the coupled weight decay is $10^{-5}$, and the global gradient norm is clipped at 5.

The first code in each cached pair is drawn from four sources: a real training code, a corrupted training code, a uniform random code, or a teacher trajectory, with probabilities $0.35$, $0.25$, $0.20$, and $0.20$, respectively. The corrupted source flips each spin independently using a per-code probability drawn uniformly from $0.05$ to $0.5$. The second code is obtained by perturbing the first with a per-pair flip probability sampled uniformly from $1/32$ to $0.5$; if none flips, a single flip is forced. Teacher trajectories use 64 chains with cycling class labels and 16 single-spin Metropolis sweeps on a linear temperature schedule from 2 to 1; configurations are saved after sweeps 4, 8, 12, and 16. We evaluate both codes at $c=0$ under every candidate label. The training and validation cache seeds are 2026 and 102029, respectively. These are procedures for constructing the targets, distinct from the QPU and logical-graph SQA samplers used for refinement.

\section{Classical Inference and Image Restoration}
\label{app:inference}
Classification scores each of the ten candidate labels on the same code. The teacher minimizes $E_{\rm H}$ with $c=0$, while each student minimizes $F_\theta$. Restoration starts its first trajectory from the code of the observed image and its second from a copy with each latent spin independently flipped with probability 0.1. The initial label is the energy-based prediction for that trajectory's starting code. Joint SA initializes the auxiliary spins independently at random. We measure fidelity to the available pixels by
\begin{equation}
 \mathcal D_{\rm obs}(z)=
 \frac{\sum_i m_i[D(z)_i-x_i^{\rm obs}]^2}{\sum_i m_i}.
\end{equation}
Here $i$ indexes pixels, $D(z)_i$ is pixel $i$ of the decoded candidate, and $x_i^{\rm obs}$ is the input value at that pixel. The mask $m_i$ equals one on observed pixels and zero elsewhere; the denominator counts the observed pixels, so $\mathcal D_{\rm obs}$ is their mean squared discrepancy. The visible sampler (Visible + 32 sweeps in Fig.~\ref{fig:inference}) uses the score $F_\theta(z;y)+\lambda\mathcal D_{\rm obs}(z)$, so the auxiliary spins enter through Eq.~\eqref{eq:free}. Joint SA (simulated annealing) uses $E_\theta(z,a;y)+\lambda\mathcal D_{\rm obs}(z)$ and keeps all auxiliary spins explicit. The coefficient $\lambda$ sets the weight of agreement with the observation relative to the learned energy. The algorithms propose spin flips with a Metropolis acceptance rule \cite{metropolis} and draw the categorical label from its conditional distribution. The annealing stage decreases the temperature over successive sweeps \cite{kirkpatrick}. A sweep visits the active spin variables; the two state spaces have different numbers of active variables. The protocols use 32 initial sweeps with temperatures from 2 to 1, then 32 additional sweeps at the terminal temperature. During cooling, the visible sampler tempers the fixed $T=1$ free energy, whereas the nominal visible marginal of the joint sampler follows the free energy at the current temperature (Appendix~\ref{app:representation}). The nominal visible targets agree at $T=1$. The reported outputs are from the end of the additional sweeps.

Both methods use $\lambda=1000$ from the preceding validation protocol, the same masks and observations, and two trajectories per image retained in the metrics. The five conditions are clean inputs, Gaussian noise with standard deviation $0.15$ or $0.30$ in unit pixel range, and masking of $25\%$ or $50\%$ of pixels. Reconstruction MSE is evaluated against the full clean image. Ground-truth labels are used to score the sampled label and the CNN prediction on the decoded image. The direct autoencoder baseline encodes and decodes the observed image once.

\section{Sampling and the Refinement Gradient}
\label{app:sampling}
For the ideal conditional model at logical $T=1$, the probability of a visible code is
\begin{equation}
 p_\theta(z\mid y)=\frac{\exp[-F_\theta(z;y)]}{Z_\theta(y)},
 \qquad Z_\theta(y)=\sum_w\exp[-F_\theta(w;y)].
\end{equation}
Here $Z_\theta(y)$ is the normalizing sum over all possible visible codes $w$ at fixed class. The gradient of the negative log probability of a data code is
\begin{equation}
 \nabla_\theta[-\log p_\theta(z\mid y)] =
 \nabla_\theta F_\theta(z;y)-
 \E_{w\sim p_\theta(\cdot\mid y)}[\nabla_\theta F_\theta(w;y)] .
\end{equation}
The operator $\nabla_\theta$ differentiates with respect to the student coefficients. The first term is the free-energy gradient at a data code $z$; the second is its expectation over codes $w$ generated by the conditional model. Equation~\eqref{eq:surrogate} averages the data term over the positive batch and replaces the expectation by the negative-batch mean, yielding an update that depends on the sampler. Both gradients marginalize the auxiliaries analytically; the negative phase receives the visible part of each joint read. Fixed offsets within a requested class cancel in the phase difference. The 100-step refinement uses a fresh Adam state with learning rate $10^{-3}$, $(\beta_1,\beta_2)=(0.9,0.999)$, $\epsilon=10^{-8}$ and global clipping at 5. Per paired starting checkpoint the two arms share the positive batches, class visits and optimizer settings. QPU evaluation contains three blocks, each with 100 reads for each class and gauge, using two gauges. The main hardware protocol uses raw readout, disables automatic rescaling, enables intersample-correlation reduction, and specifies programming and readout thermalization times of $1{,}000\,\mu\mathrm{s}$ and $0\,\mu\mathrm{s}$, respectively. Logical-graph SQA evaluation uses three sampling streams and 200 images per class and stream.

For the logical-graph SQA protocol, each read starts from a uniform binary configuration of the 160 logical spins, copied across its 16 Trotter slices. The transverse field follows $\Gamma_k=4(0.05/4)^{k/63}$, where $k=0,\ldots,63$ is the sweep index. The implementation permutes the 32 single-visible-spin groups and the auxiliary group, updates even and odd slices, and proposes full Trotter-line flips. One slice is selected uniformly at readout. Logical-graph SQA evaluation uses streams seeded at 620261, 620262, and 620263.

The hardware compiler distributes a logical field uniformly across its chain and a logical pairwise coupling uniformly across the available contacts between the corresponding chains. Each gauge sign is propagated to all qubits in a logical chain. The evaluation uses the identity gauge and a second gauge that reverses 81 logical variables.  The mapping and the gauge vectors used in these experiments are retained in the run records.

For embedded-graph SQA, Eq.~\eqref{eq:physical_anneal} uses the standard annealing schedules for Advantage2\_system1, tabulated in GHz \cite{dwaveschedules}. At 20\,mK, $k_{\rm B}T/h_{\rm P}=0.41673$\,GHz. With $P$ Trotter slices, an adjacent-slice ferromagnetic coupling in the dimensionless action has magnitude $\frac12\log\coth[h_{\rm P}A(s)/(2P k_{\rm B}T)]$, and the coefficient of $H_{\rm prog}$ in the diagonal Boltzmann exponent is $-h_{\rm P}B(s)/(2Pk_{\rm B}T)$ \cite{suzuki,martonak}. The anneal schedule is sampled along 64 Monte Carlo sweeps, with independent random starts for each update and readout from a randomly chosen slice. We then undo the gauge and apply the same majority rule as on hardware. The 20\,mK value is the simulation bath parameter, while the 500\,$\mu$s value specifies the physical QPU schedule. In the physically initialized pilot the baseline and final evaluations were acquired in separate sessions, before and after the refinement, respectively.

\section{Generation Metrics}
\label{app:metrics}
The conditional accuracy and distinct-code fraction are
\begin{equation}
 \mathrm{Acc}=\frac{1}{n}\sum_{i=1}^n\ind[g(x_i)=y_i],\qquad
 f_{\rm distinct}=\frac{|\{z_i:i=1,\ldots,n\}|}{n}.
\end{equation}
Here $n$ is the number of draws, $z_i$ is the sampled visible code, $x_i=D(z_i)$ is its decoded image, $y_i$ is its requested label, and $g$ is the frozen convolutional neural network (CNN) classifier. The indicator is one when the predicted class equals the requested class and zero otherwise. The set in the second numerator counts each distinct code once. Both denominators retain repeated draws.

For real-image recall, a generated point in the 64-dimensional CNN feature space is assigned the Euclidean distance to its fifth nearest generated neighbor from the same requested class (excluding the point itself). The union of these balls estimates the covered portion of feature space for that class \cite{knn}. We report the fraction of the 200 held-out real reference images for that class in this union, including all generated samples of the requested class when constructing the balls. Macro recall is the arithmetic mean over the ten class values. The reconstruction-reference version uses the autoencoder reconstructions of these same held-out images.

The Fr\'echet score adapts the Gaussian feature statistic of Ref.~\cite{fid} to the same frozen CNN features:
\begin{equation}
 \|\mu_g-\mu_r\|^2+
 \operatorname{tr}\left(C_g+C_r-2(C_r^{1/2}C_gC_r^{1/2})^{1/2}\right).
\end{equation}
Here $\mu_g,C_g$ are the mean and covariance of generated-image features, and $\mu_r,C_r$ are those of the reference-image features. The squared norm measures the separation of the means; the trace term measures the covariance discrepancy, with matrix square roots taken in the positive-semidefinite sense. We compute the score separately for the real and reconstruction references.

\section{Paired Refinement Effects}
\label{app:paired}
Table~\ref{tab:paired} gives the changes for the individual starting checkpoints in Fig.~\ref{fig:cross}. Each difference subtracts the frozen student's metric from the refined student's metric with the same evaluation sampler, after averaging acquisition blocks or simulation streams within the checkpoint.
\begin{table}[!htbp]\centering
\small
\caption{Within-seed change after refinement, in percentage points. Recall uses real images of the requested class as reference.}
\label{tab:paired}
\begin{tabular}{@{}rllrr@{}}
\toprule
Seed & Evaluation & Refinement & $\Delta$ accuracy & $\Delta$ recall \\
\midrule
721201 & QPU & QPU & +29.03 & +19.20 \\
721201 & QPU & Logical-graph SQA & -1.82 & +8.87 \\
721201 & Logical-graph SQA & QPU & -7.53 & -1.02 \\
721201 & Logical-graph SQA & Logical-graph SQA & -1.70 & +19.22 \\
721202 & QPU & QPU & +7.43 & +21.87 \\
721202 & QPU & Logical-graph SQA & -4.17 & +13.88 \\
721202 & Logical-graph SQA & QPU & -4.13 & -1.67 \\
721202 & Logical-graph SQA & Logical-graph SQA & -2.85 & +12.82 \\
721203 & QPU & QPU & +25.58 & +21.50 \\
721203 & QPU & Logical-graph SQA & +5.85 & +7.93 \\
721203 & Logical-graph SQA & QPU & -10.68 & +3.15 \\
721203 & Logical-graph SQA & Logical-graph SQA & -2.78 & +20.47 \\
\bottomrule
\end{tabular}
\end{table}

\section{Logged Objective During Hardware Refinement}
\label{app:surrogate}
\begin{figure}[!t]\centering
\includegraphics[width=.89\linewidth]{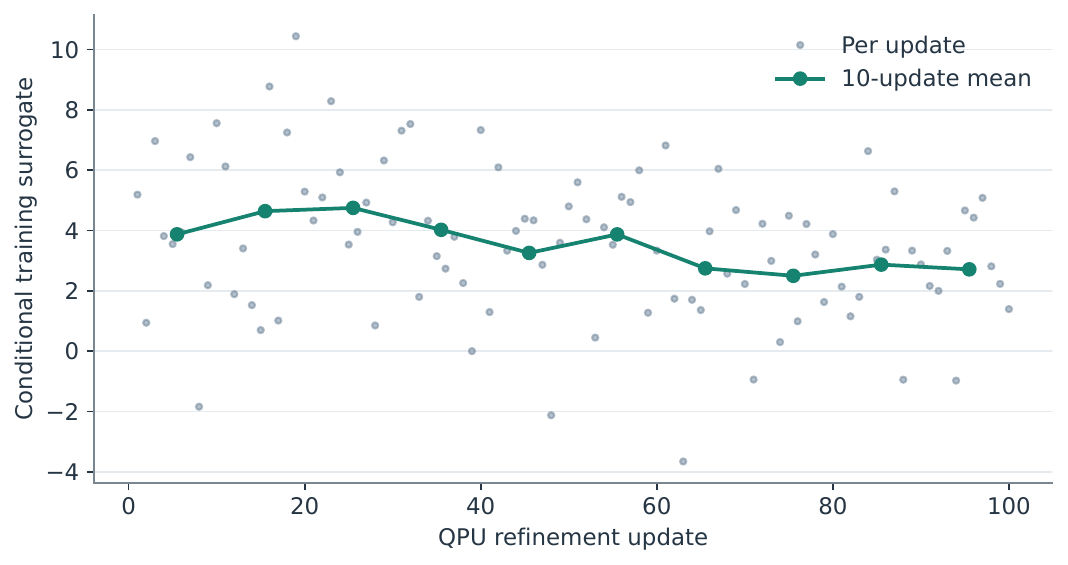}
\caption{\textbf{Conditional training surrogate.} Grey points are the difference of data and sample mean free energies for the current class and batches. Green points are the averages over successive ten-update cycles.}
\label{fig:surrogate}
\end{figure}

Figure~\ref{fig:surrogate} gives the value of Eq.~\eqref{eq:surrogate} logged at each update in Fig.~\ref{fig:trajectory}. A positive value means the observed QPU codes have lower mean free energy than the data codes for the class scheduled at that step. The smoothed series averages successive nonoverlapping cycles of ten updates, each of which visits every class. The recorded batches vary with the iterates and scheduled class; generation accuracy is evaluated separately in Fig.~\ref{fig:trajectory}.

\section{Restricted Resampling of Broken Chains}
\label{app:rrs}
\begin{figure}[!t]\centering
\includegraphics[width=.90\linewidth]{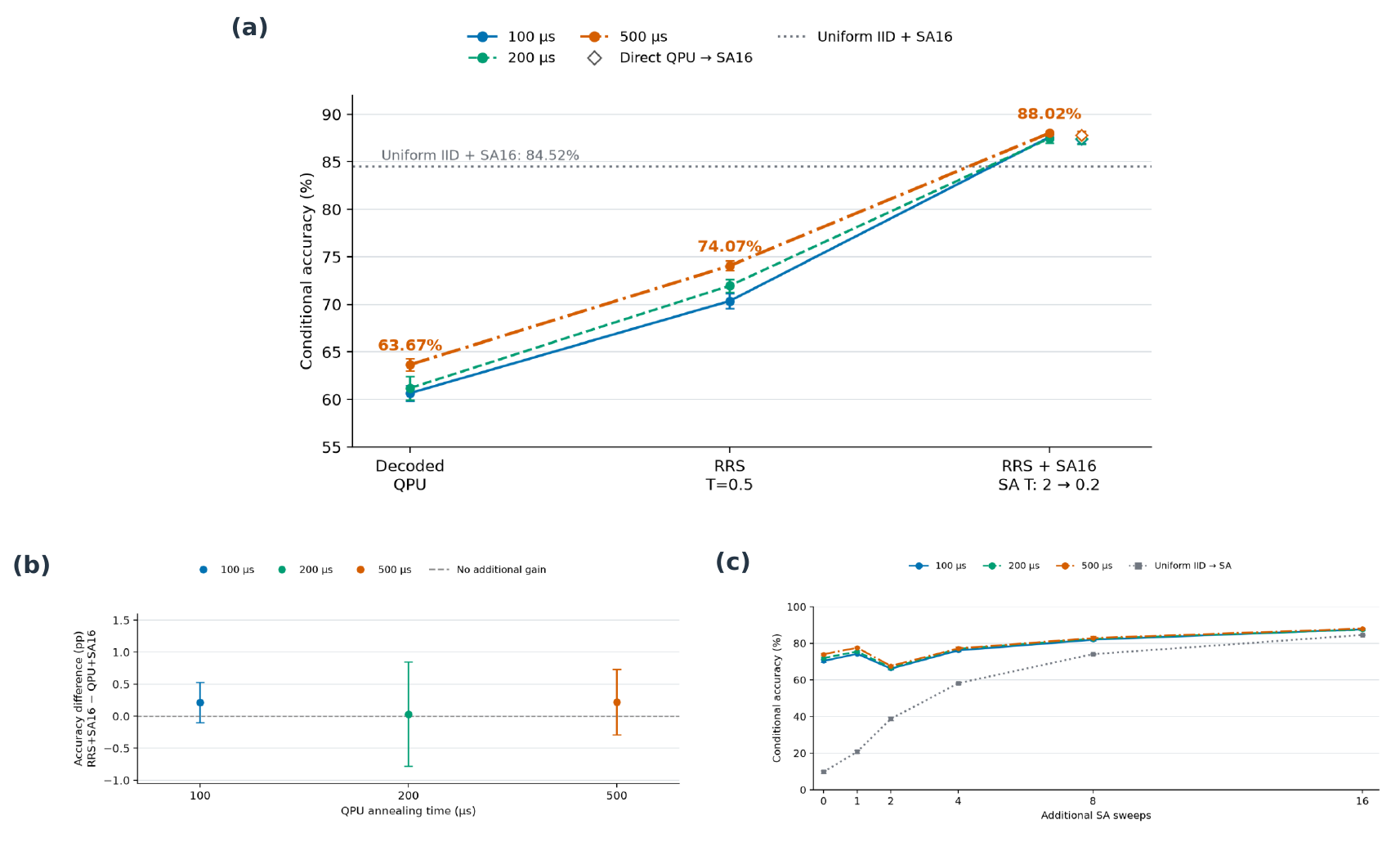}
\caption{\textbf{Restricted resampling and classical postprocessing.} (a) Conditional accuracy of majority-vote QPU decoding, RRS, and RRS followed by 16 SA sweeps at the indicated anneal times. Diamonds denote direct QPU decoding followed by the same SA16 protocol. The dotted reference is uniform random initialization followed by SA16. (b) Paired accuracy difference from inserting RRS before SA16 on matched reads. Error bars in (a,b) describe the spread across three acquisition blocks after averaging nested repeats where applicable. (c) Accuracy after different SA budgets from QPU reads completed by RRS and from uniform random starts. The zero-sweep points are the initial states. Each nonzero sweep budget uses its own complete temperature schedule. Error bars show the sample standard deviation across three blocks after averaging nested realizations.}
\label{fig:rrs}
\end{figure}

Restricted resampling (RRS) follows the conditional resampling approach for broken chains in Ref.~\cite{marshall}. The conditional law used to fill in the broken chains is
\begin{equation}
 p(s_B\mid s_{\bar B},y)=
 \frac{\exp[-E_\theta(s_B,s_{\bar B};y)/T_{\rm RRS}]}
 {\sum_{s'_B}\exp[-E_\theta(s'_B,s_{\bar B};y)/T_{\rm RRS}]} .
\end{equation}
Here $B$ is the set of logical variables whose physical chain qubits disagree, $s_B$ is a candidate assignment to them, and $s_{\bar B}$ holds the logical values of the intact chains fixed. The denominator normalizes over all candidate assignments $s'_B$ at the same fixed class $y$. We enumerate these assignments exactly and use $T_{\rm RRS}=0.5$ in logical units.

Figure~\ref{fig:rrs} uses an archived frozen dense student with uniform assignment of logical coefficients over physical chain elements or contacts. At 100, 200, and 500\,$\mu$s, respectively, RRS raised accuracy from 60.67, 61.20, and 63.67\% to 70.36, 71.97, and 74.07\% before any classical sweeps. With 16 subsequent SA sweeps, the mean paired increment attributable to adding RRS before SA was 0.22, 0.03, and 0.22 percentage points, respectively. Thus chain decoding affects the class accuracy of the QPU outputs before polishing; after 16 sweeps the two decoding choices give similar accuracy in this cohort.

\section{Hardware Sampling and Postprocessing Across Student Graphs}
\label{app:architecture}
\begin{figure}[!tb]\centering
\includegraphics[width=.90\linewidth]{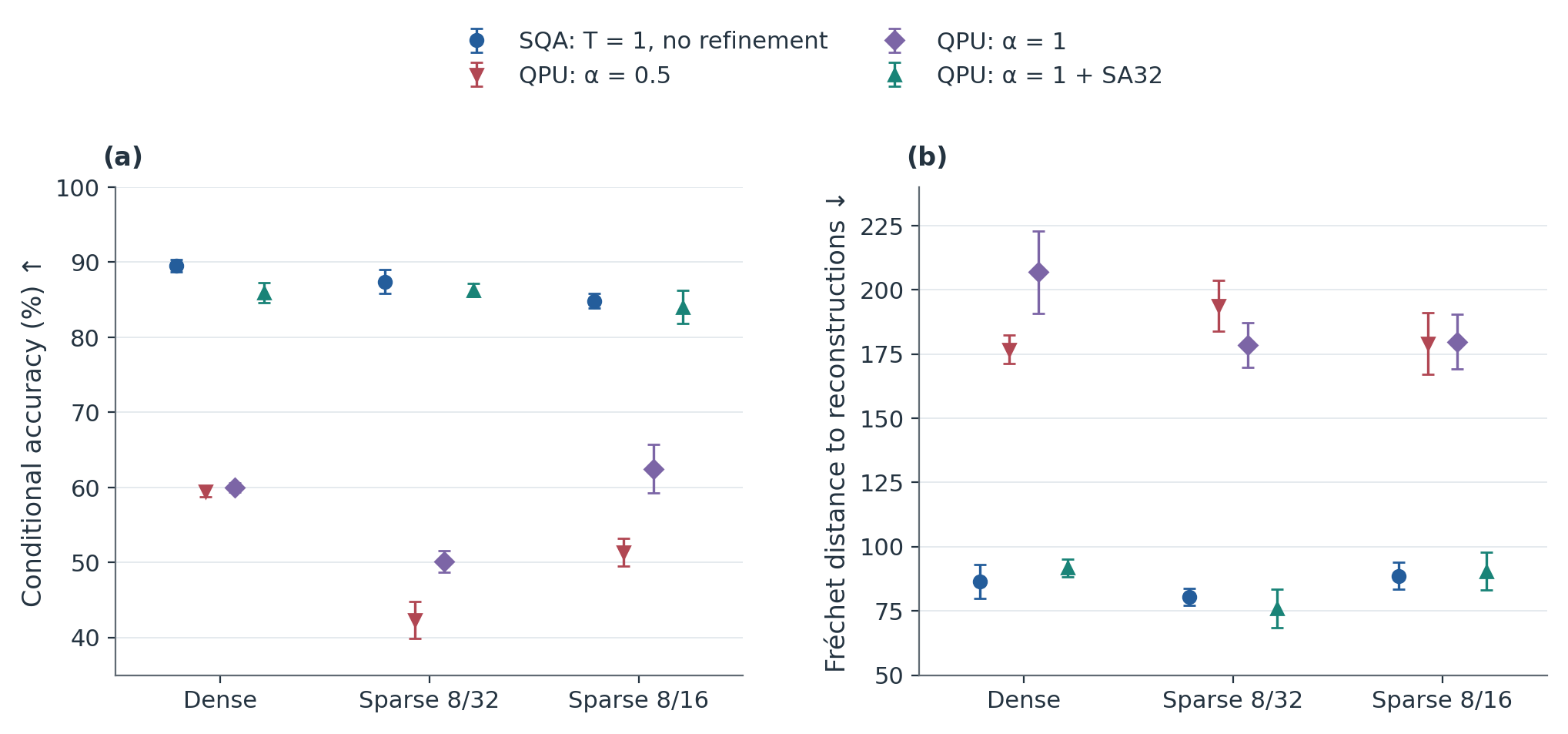}
\caption{\textbf{Class-conditioned generation across student graphs.} (a) Conditional accuracy and (b) Fr\'echet distance to autoencoder reconstructions in fixed CNN feature space. Each architecture is evaluated with logical-graph SQA, two hardware programming fractions, and 32 fixed-temperature Metropolis sweeps (labeled SA32 in the archived plots) applied to hardware reads at $\alpha=1$. Error bars denote the sample standard deviation across five hardware acquisitions or five logical-graph SQA sampling seeds at fixed checkpoints.}
\label{fig:architectures}
\end{figure}

\begin{figure}[!htbp]\centering
\includegraphics[width=.92\linewidth]{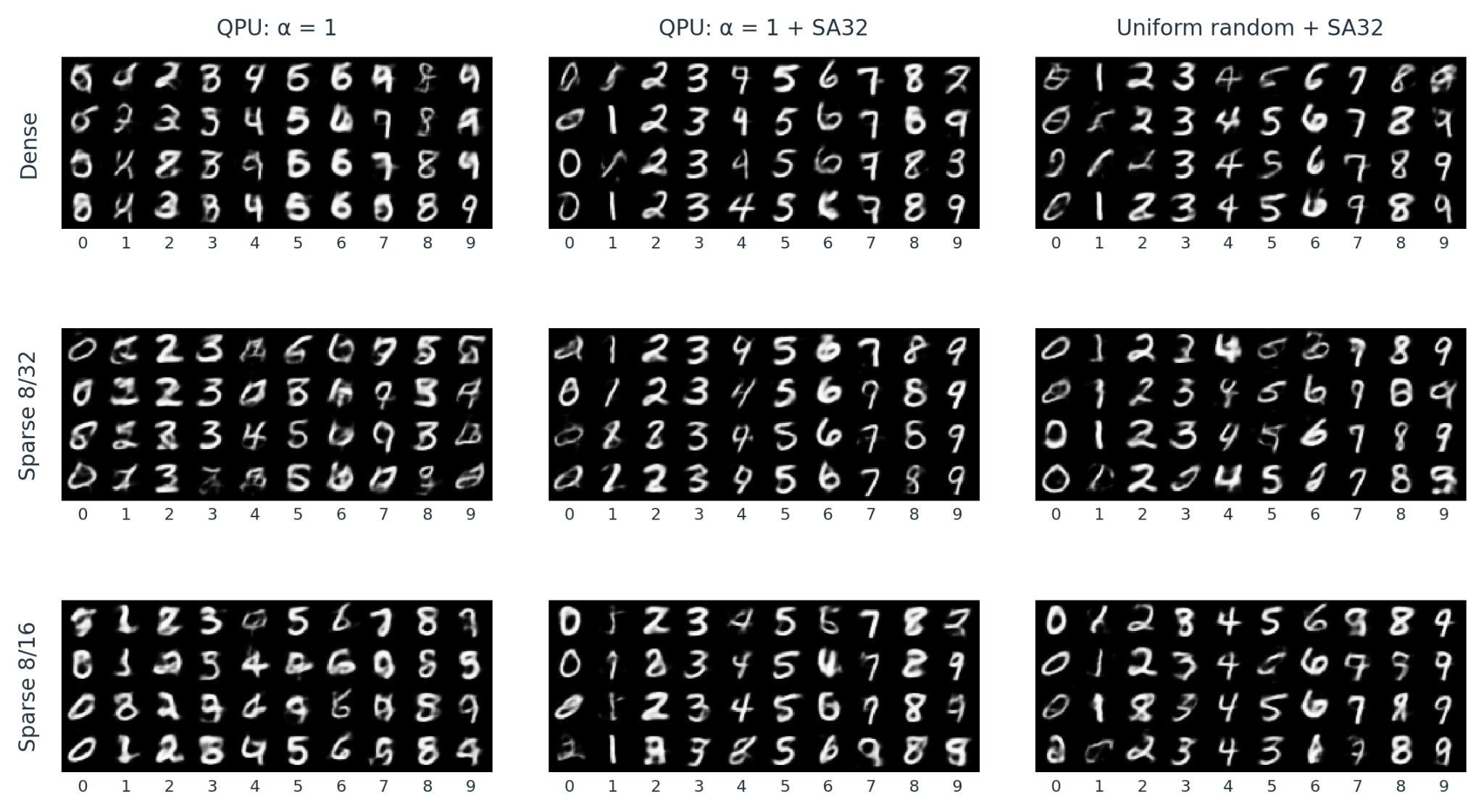}
\caption{\textbf{Samples before and after classical postprocessing.} SA32 denotes 32 fixed-temperature Metropolis sweeps. Rows identify the three student architectures. The three groups show QPU reads at $\alpha=1$, QPU reads followed by SA32, and uniformly initialized states followed by SA32. Each group shows the first four archived examples per requested digit. The energy parameters remain fixed throughout the sampling procedures.}
\label{fig:architecture_digits}
\end{figure}

We evaluated the three archived student architectures from Fig.~\ref{fig:inference} on a separate hardware campaign. Sparse 8/32 preserves dense visible-auxiliary connectivity with a degree-eight visible graph; Sparse 8/16 also reduces the number of visible neighbors per auxiliary spin. The programmed energy fraction $\alpha$ denotes the selected fraction of the reference programming scale in that campaign. We compare logical-graph SQA at $T=1$, majority-decoded QPU reads at $\alpha=0.5$ or $1$, and QPU reads at $\alpha=1$ followed by 32 joint Metropolis sweeps at fixed logical temperature $T=1$ and fixed weights (Metropolis-32). The hardware campaign used a $20\,\mu$s forward anneal and model-specific embeddings with 1,064, 1,074 and 874 physical qubits for Dense, Sparse 8/32 and Sparse 8/16, respectively. At $\alpha=1$ the programmed coefficient scales were $0.75563$, $0.51857$ and $0.43597$. Each condition includes five repeats of 500 generated images. Fr\'echet distances use 2,000 autoencoder reconstructions of validation images as the reference.

Figure~\ref{fig:architectures} shows that postprocessing changed both conditional accuracy and the distance to reconstructions. At $\alpha=1$, the raw QPU accuracies were $60.00\%$, $50.16\%$ and $62.48\%$ for Dense, Sparse 8/32 and Sparse 8/16; the corresponding GPU Metropolis-32 outputs reached $85.92\%$, $86.28\%$ and $84.04\%$. The dense logical-graph SQA result was $89.56\%$. These are sampling outcomes for fixed trained models.

Figure~\ref{fig:architecture_digits} also shows outputs initialized uniformly at random and run through Metropolis-32. In the matched CPU postprocessing study, the dense student gave $86.68\pm0.77\%$ from $\alpha=1$ QPU reads and $88.28\pm1.21\%$ from uniform random starts. Uniform-start Metropolis-32 exceeded QPU-start Metropolis-32 by 1.60 percentage points on average across the five repeats. The CPU control and the GPU summary in Fig.~\ref{fig:architectures} are reported separately.

\section{Training from Random Initialization with QPU Negative Samples}
\label{app:qpu-from-scratch}

We initialized the coefficients of the Dense model (32 visible and
128 auxiliary spins) at random and trained them for 8{,}570 updates
using the conditional free-energy objective of
Appendix~\ref{app:sampling}. For each update, positive examples came
from encoded MNIST training images, and negative examples came from
the QPU programmed with the current coefficients and the same class.
The energy model was trained without teacher targets or SQA
pretraining. The autoencoder and the CNN used to evaluate generated
images remained fixed. The single training run used 862.15\,s of
charged QPU time; the final evaluation used a further 6.03\,s.

We froze the final coefficients and acquired 6{,}000 fresh QPU reads
in three blocks of 2{,}000 images, balanced across the ten requested
digits. We decoded chains by majority vote after undoing the gauge.
For the classical postprocessing comparison, we initialized
simulated annealing (SA) from these same decoded joint spin states
and held the coefficients and class fixed. Each sweep proposed
single-spin Metropolis updates on the visible and auxiliary spins.
Each sweep budget used its own full geometric cooling schedule from
logical temperature 2 to 0.2. We ran three SA realizations per
acquisition block and averaged their metrics within that block.

Conditional accuracy was 73.27\% for the raw QPU reads. It was
66.06\%, 78.30\%, and 84.81\% after 4, 8, and 16 SA sweeps,
respectively. The 16-sweep setting also reduced the Fr\'echet
distance to original images from 222.03 to 211.21. Over the same
comparison, macro recall relative to the original images decreased
from 52.60\% to 36.91\%, and the fraction of distinct visible
codes decreased from 99.97\% to 91.38\%
(Fig.~\ref{fig:qpu-from-scratch}; see Appendix~\ref{app:metrics}).
For this trained model, the 16-sweep postprocessing improved
agreement with the requested class while reducing coverage and
latent diversity. The postprocessing required no additional QPU
reads or parameter updates.

\begin{figure}[!htbp]
   \centering
   \includegraphics[width=\linewidth]{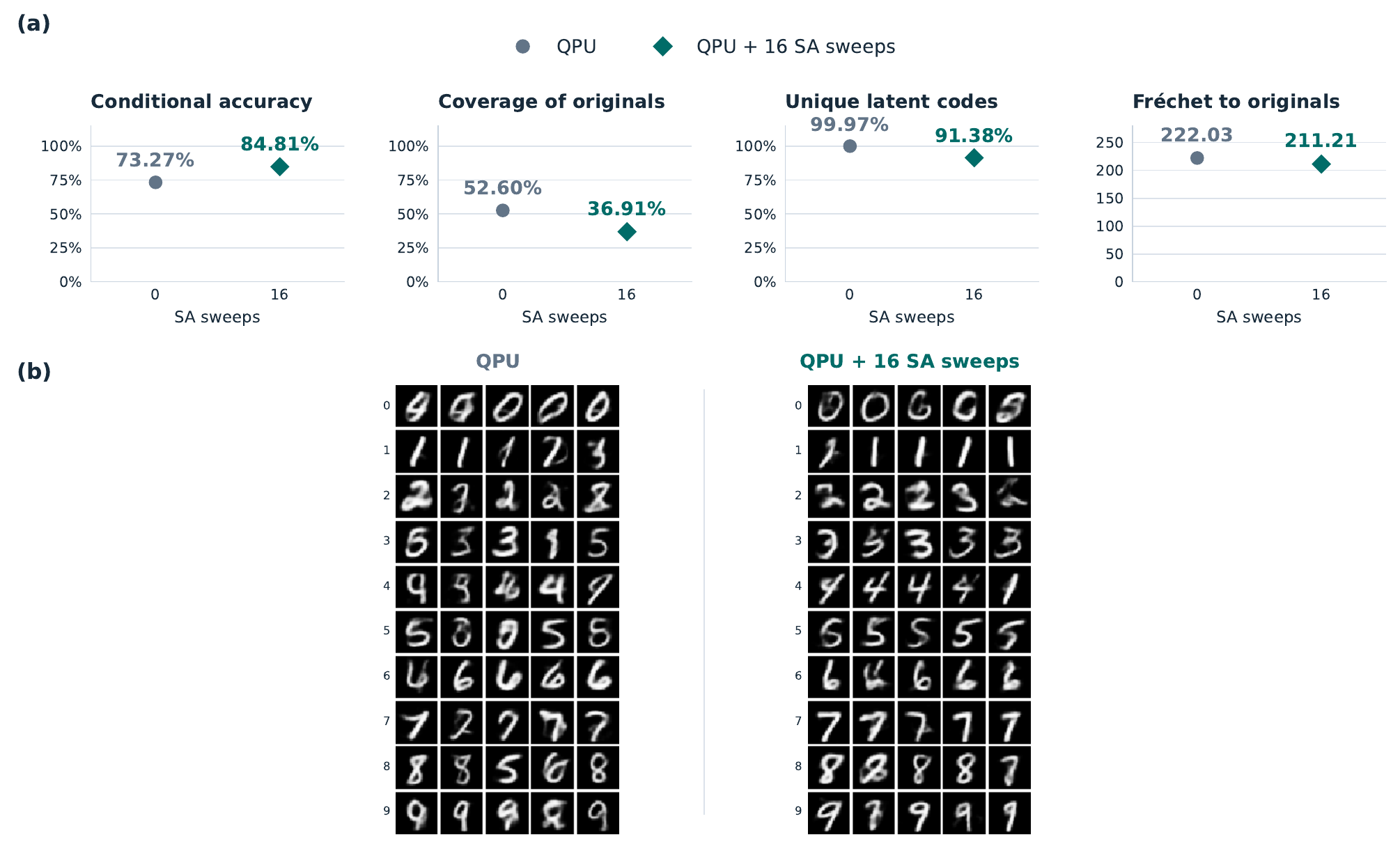}
   \caption{\textbf{Generation after training from random initialization
   with QPU negative samples.} (a) Conditional accuracy, macro recall
   relative to original images, fraction of distinct visible codes, and
   Fr\'echet distance to original images for raw QPU reads and the same
   reads followed by 16 SA sweeps. Error bars show the sample standard
   deviation across three acquisition blocks at one trained checkpoint.
   (b) Five paired outputs per requested digit from the first block and
   the first gauge, before and after SA. The examples follow acquisition
   order and were not filtered by quality.}
   \label{fig:qpu-from-scratch}
\end{figure}

\clearpage
\section{Binary Autoencoder}
\label{app:binary-ae}
The frozen autoencoder maps a $28\times28$ grayscale image to a 32-bit representation. The encoder uses two padded $3\times3$ convolutions with stride 2 and channel widths 32 and 64. Each convolution is followed by group normalization with eight groups and a SiLU activation \cite{groupnorm,silu}. The resulting $64\times7\times7$ activations are flattened and projected to 32 logits $r_i$. The binary code has entries $b_i=\ind[r_i\geq 0]$; the backward pass uses the identity straight-through gradient \cite{bengioSTE}. The decoder projects the binary vector to $64\times7\times7$, applies a stride-one convolutional block, and upsamples twice by a factor of two using nearest neighbors. A convolutional block between the two upsampling steps reduces the width from 64 to 32 channels. A final padded $3\times3$ convolution and sigmoid produce one output channel. The energy model uses $z=2b-1$, and $D(z)$ reconstructs the image from $(z+1)/2$.

The training objective combines pixel mean squared error, a commitment term with weight $0.25$, and a logit penalty with weight $10^{-4}$. The commitment term is the mean squared difference between the sigmoid of the logits and the detached hard binary codes, and the penalty is the mean squared logit. Training used Adam \cite{adam} for 30 epochs with batch size 64, learning rate $10^{-3}$ and training seed 0. The cache manifest records a split of 50,000 training, 10,000 validation, and 10,000 test images, with split seed 2026. We use the checkpoint at epoch 28, selected by the lowest validation pixel mean squared error, and keep the encoder and decoder fixed in every reported energy-model experiment.
\end{document}